\documentclass[sigplan,nonacm]{acmart}

\usepackage{graphicx}
\usepackage{url}
\usepackage{amsmath}
\usepackage{xcolor}
\usepackage[linesnumbered,ruled,vlined]{algorithm2e}
\usepackage{siunitx}
\usepackage{tabularx}
\usepackage{listings}
\usepackage{float}

\newcommand{\secref}[1]{Sec.~\ref{#1}}
\newcommand{\figref}[1]{Fig.~\ref{#1}}
\newcommand{\tabref}[1]{Table~\ref{#1}}

\begin{document}

\title{GreenPeas: Unlocking adaptive quantum error correction with just-in-time decoding hypergraphs}

\begin{abstract}
  Circuit-level decoders are essential for the realisation of low-overhead fault-tolerant quantum computing.
  However, they rely on complex hypergraphs that are traditionally compiled ahead-of-time.
  This static approach introduces a significant bottleneck for an emerging class of adaptive circuits, where the structure is modified during execution based on mid-circuit measurement outcomes.
  Pre-compiling hypergraphs for all possible circuit branches would incur an exponential memory cost, rendering current tools impractical for these workloads.
  Hence, we introduce GreenPeas, a just-in-time compiler for decoding hypergraphs.
  By lowering the realised circuit to a space-time error propagation graph, GreenPeas decomposes Stim's backtracking algorithm for error analysis into two sequentially dependent, internally parallelisable stages: (1)~mapping physical errors to their corresponding equivalence classes, and (2)~aggregating error probabilities within each class.
  Evaluated on surface and bivariate bicycle code memory circuits without user-annotated repeat blocks, GreenPeas achieves a geometric mean speedup of $13.2\times$ over Stim using a high-end GPU.
  This speedup carries over to the adaptive regime, unlocking circuit-level decoding of $[[4,2,2]]$-concatenated surface code memories with adaptive syndrome measurements---a capability previously restricted to less accurate phenomenological decoders---yielding $6.7\times$ lower logical error rate and $4.5\times$ lower decoding latency at a representative outer code distance of 10.
\end{abstract}

\author{Abbas B. Ziad}
\email{abbas.ziad25@imperial.ac.uk}
\orcid{0009-0008-2517-0619}
\affiliation{%
  \institution{Imperial College London}
  \city{London}
  \country{UK}
}

\author{Jubo Xu}
\email{jubo.xu20@imperial.ac.uk}
\orcid{0009-0009-6314-5833}
\affiliation{%
  \institution{Imperial College London}
  \city{London}
  \country{UK}
}

\author{Hongxiang Fan}
\email{hongxiang.fan@imperial.ac.uk}
\orcid{0000-0003-2387-5611}
\affiliation{%
  \institution{Imperial College London}
  \city{London}
  \country{UK}
}

\maketitle

\section{Introduction}
\label{sec:introduction}

The theoretical potential of quantum computing promises to revolutionise fields such as cryptography~\cite{shor_polynomial-time_1997}, material science~\cite{lloyd_universal_1996} and drug discovery~\cite{aspuru-guzik_simulated_2005}.
However, realising these applications in practice requires error rates as low as $10^{-12}$ to support the trillions of operations involved~\cite{cain_shors_2026}.
Current devices remain far from this regime, with fidelities of "three nines" ($99.9\%$) considered state-of-the-art~\cite{google_quantum_ai_and_collaborators_quantum_2025}.
This three-to-nine order-of-magnitude gap necessitates robust quantum error correction to bridge the divide between noisy physical and reliable logical qubits.

In general, a quantum error correction system consists of two core components: (a)~a quantum code that encodes many noisy physical qubits into one or more reliable logical qubits, and (b)~a classical decoder that infers logical errors by processing syndromes extracted from the code via syndrome measurement (SM) circuits.
These decoders typically operate under one of two paradigms: phenomenological, which uses a simplified model of data qubit and measurement errors, or circuit-level, which employs a more complex hypergraph to capture all error mechanisms in the SM circuit.

While circuit-level decoders offer higher accuracy than their phenomenological counterparts, mapping an SM circuit under circuit-level noise onto a decoding hypergraph incurs a substantially higher computational overhead.
Even with state-of-the-art tools like Stim~\cite{gidney_stim_2021}, this translation takes up to ten milliseconds per SM cycle, making it ill-suited for online compilation, even on "slow clock" architectures with cycle times around 1 millisecond.
This latency also inflates simulation times, rendering high-confidence fidelity estimation impractical for such online use cases~\cite{berthusen_adaptive_2025}.

Alternative offline approaches impose a critical bottleneck for adaptive SM circuits~\cite{tansuwannont_adaptive_2023,bhatnagar_low-depth_2023,berthusen_adaptive_2025,paszko_dynamic_2025,biswas_noise-adapted_2024}, where gate sequences are determined at runtime based on mid-circuit measurement outcomes.
Since this paradigm necessitates pre-computing all possible hypergraphs---a task that scales exponentially in the number of circuit branches---this places an unsustainable burden on the system. 
Similar challenges appear in logical circuits with conditional S-gate corrections~\cite{bravyi_universal_2005,fowler_surface_2012}.
Offline compilation thus relegates circuit-level decoders to static circuits or adaptive circuits with a low branching factor.

To overcome this limitation, we introduce GreenPeas\footnote{\url{https://github.com/abbrazi/GreenPeas}}, an online, just-in-time compiler for decoding hypergraphs with ultra-low, sub-millisecond per-cycle latency.

Our contributions are as follows:

\begin{itemize}
    \item \textbf{Parallel error analysis:} We decompose Stim’s backtracking algorithm into two sequentially dependent, internally parallelisable stages: (1)~mapping physical errors to their corresponding equivalence classes, and (2)~aggregating error probabilities within each class using space-time error propagation graphs.
    \item \textbf{Over an order-of-magnitude latency reduction:} We achieve an average $13.2\times$ speedup over Stim on a high-end GPU backend across surface and bivariate bicycle code memory circuits without user-annotated repeat blocks, which serve as worst-case proxy workloads for round-by-round adaptive circuits.
    \newpage  
    \item \textbf{Circuit-level decoding of adaptive SM circuits:} We demonstrate that GreenPeas enables circuit-level decoding of adaptive syndrome measurements on the $[[4,2,2]]$-concatenated surface code, yielding up to $6.7\times$ fewer logical errors and $4.5\times$ faster decoding when compared to prior phenomenological methods.
    \item \textbf{Configurable correlation level:} We introduce a multi-level hyperparameter that moves beyond the conventional uncorrelated vs.\ correlated dichotomy, enabling users to finely navigate the accuracy-speed trade-off during hypergraph compilation.
\end{itemize}

\section{Background, motivation and formalism}
\label{sec:background-motivation-and-formalism}

This section reviews the core concepts involved in decoding hypergraph compilation, motivates the need for low-latency online compilation, and presents the decoupled formulation of backtracking that underpins our work.

\subsection{Background}
\label{sec:background-motivation-and-formalism:background}

\subsubsection{Stabiliser codes}
\label{sec:background-motivation-and-formalism:background:stabiliser-codes}

An $[[n,k,d]]$ stabiliser code encodes $k$ logical qubits into $n$ physical qubits with distance $d$, the minimum number of physical errors that can cause an undetectable logical error.
It is defined by stabiliser generators whose measurement yields the error syndrome~\cite{gottesman_stabilizer_1997}.

\subsubsection{Syndrome measurement circuit}
\label{sec:background-motivation-and-formalism:background:syndrome-measurement-circuit}

In practice, the stabiliser generators of a code are measured via a syndrome measurement (SM) circuit, which consists of a sequence of Clifford gates, resets and measurements, as illustrated in \figref{fig:error-analysis} for the $d = 3$ bit-flip code.
Due to measurement errors, this circuit must be repeated $d$ times, motivating the concept of a detector: a parity of time-adjacent measurements of the same stabiliser generator.

\subsubsection{Noise models}
\label{sec:background-motivation-and-formalism:background:noise-models}

A pure SM circuit describes an ideal scenario where Clifford gates, resets and measurements are executed without any injected error channels.
Modelling faulty hardware requires augmenting the circuit with explicit error channels.
We describe two noise models, in order of increasing realism: phenomenological and circuit-level.

\paragraph{Phenomenological}
\label{sec:background-motivation-and-formalism:background:noise-models:phenomenological}

The phenomenological noise model introduces error channels on data qubits and measurements while leaving Clifford gates and resets fault-free. 
Before each round, the data qubits are subject to a depolarising error channel with probability $p_i$, and both the data and ancilla qubit measurements are flipped with probability $p_m$.

\paragraph{Circuit-level}
\label{sec:background-motivation-and-formalism:background:noise-models:circuit-level}

The circuit-level noise model introduces error channels after every gate layer. 
Specifically, single- and two-qubit Clifford gates are followed by depolarising error channels with probabilities $p_1$ and $p_2$, respectively; data and ancilla qubit measurements are flipped with probability $p_m$, resets fail with probability $p_r$, and idle qubits are subject to a depolarising error channel with probability $p_i$.

\subsubsection{Detector error model}
\label{sec:background-motivation-and-formalism:background:detector-error-model}

Under a given noise model, physical errors within an SM circuit propagate through it and trigger detectors in characteristic patterns.
Distinct errors that produce identical detector flip patterns are considered logically equivalent and are grouped into equivalence classes.
The detector error model (DEM) encodes this information, specifying the detector flip pattern and aggregate probability of each class.
Given an observed syndrome (a pattern of triggered detectors), a decoder uses the DEM to infer the most likely combination of equivalence classes to explain it.

\paragraph{Decoding hypergraph}
\label{sec:background-motivation-and-formalism:background:background:detector-error-model:decoding-hypergraph}

The DEM is often represented as a weighted hypergraph $\mathcal{G} = (\mathcal{D}, \mathcal{E})$, where each vertex $\delta \in \mathcal{D}$ denotes a detector and each hyperedge $\epsilon \in \mathcal{E}$ corresponds to an equivalence class.
The weight of a hyperedge is equal to the aggregate probability of its equivalence class.

\subsubsection{Error analysis}
\label{sec:background-motivation-and-formalism:background:error-analysis}

To construct $\mathcal{G}$, we must find the equivalence class for every possible error in the circuit, a task called error analysis.
Two strategies tackle the problem from opposite directions: forward tracking and backtracking.

\paragraph{Forward tracking}
\label{sec:background-motivation-and-formalism:background:error-analysis:forward-tracking}

\begin{figure}[htbp]
    \centering
    \includegraphics[width=\linewidth]{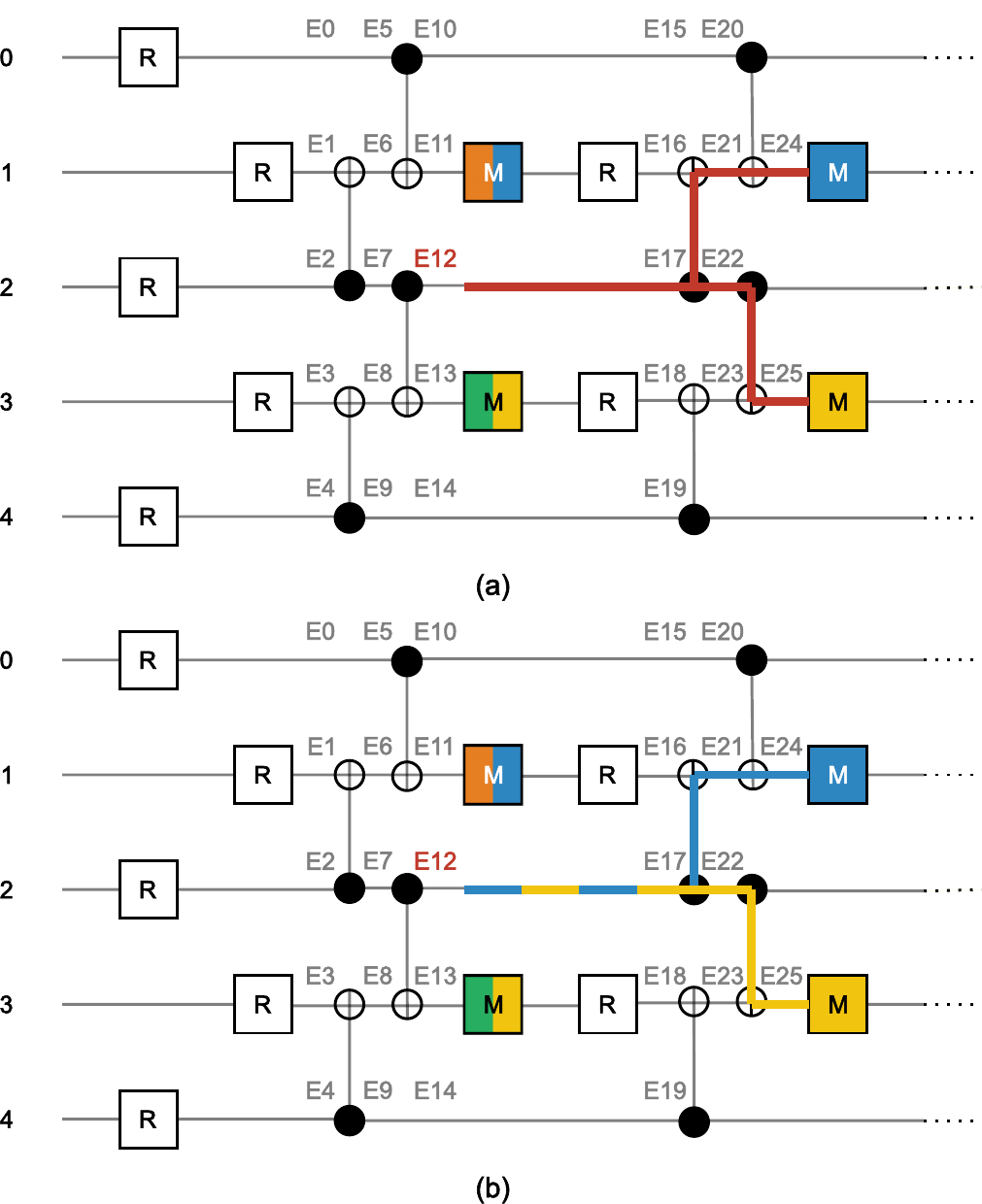}
    \caption{\textbf{Error analysis}.
    \textbf{(a)} The SM circuit for a $d = 3$ bit-flip code (two rounds).
    We show four detectors, coloured orange, blue, green and yellow.
    The orange (green) detector is defined by the first measurement of ancilla qubit 1 (3).
    The blue (yellow) detector is defined by the parity of the measurements of ancilla qubit 1 (3).
    With forward tracking, errors are tracked forward through time, e.g., the bit-flip error on data qubit 2 after the second CNOT layer ($E_{12}$) is tracked through the subsequent entangling gates to determine that it flips the yellow and blue detectors.
    \textbf{(b)} Backtracking instead tracks detector sensitivities backwards through time, e.g., the sensitivities of the yellow and blue detectors are tracked backwards from the final measurement layer, merging at the third CNOT layer to form a combined detecting region that touches $E_{12}$, revealing its equivalence class.}
    \label{fig:error-analysis}
\end{figure}

This method tracks errors forward through the circuit and records the set of detectors they flip (\figref{fig:error-analysis}(a)).
For a circuit with $n$ qubits and $l$ gate layers, it has complexity $\mathcal{O}(n^2 l^2)$~\cite{derks_designing_2025}.

\paragraph{Backtracking}
\label{sec:background-motivation-and-formalism:background:error-analysis:backtracking}

This method tracks detector sensitivities backwards through the circuit, forming detecting regions that reveal each error's equivalence class (\figref{fig:error-analysis}(b)).
It has complexity $\mathcal{O}(nl)$~\cite{gidney_stim_2021}.

\subsection{Motivation}
\label{sec:background-motivation-and-formalism:motivation}

Due to this asymptotic advantage, Stim implements backtracking rather than forward tracking.
Its algorithm, however, couples sensitivity propagation with the computation of aggregate probabilities, introducing sequential dependencies that are difficult to parallelise.
As such, per-cycle latencies reach up to ten milliseconds (see~\figref{fig:single-level}).
For even "slow clock" architectures such as neutral-atom and trapped-ion systems, whose cycle times are around 1 millisecond~\cite{bluvstein_logical_2024,tripier_fault-tolerant_2026,webster_pinnacle_2026}, this exceeds their timing budgets by an order of magnitude.

Offline compilation remains practical for static circuits with fixed noise parameters.
In long-running computations, however, physical error rates drift, so decoding hypergraphs must be recalibrated periodically to preserve logical fidelity.
Adaptive circuits pose a sharper challenge: because their structure depends on mid-circuit measurement outcomes, pre-computing a hypergraph for every execution path scales exponentially with the branching factor.
In both settings, fully offline compilation is impractical.

\subsection{Formalism}
\label{sec:background-motivation-and-formalism:formalism}

We recast backtracking as a graph processing workload with two decoupled, parallelisable stages: error equivalence class generation and probability aggregation.

\subsubsection{Space-time error propagation graph (STEPG)}
\label{sec:background-motivation-and-formalism:formalism:space-time-error-propagation-graph}

Both stages operate on the STEPG, a layered directed acyclic graph in which vertices denote potential Pauli errors and edges denote gate-induced transformations of those errors through time, e.g., an $X$ error on the control qubit before a CNOT gate spreads to an $X$ error on both the control and the target qubits after the gate (see~\figref{fig:stepg-wave}).

\subsubsection{Error equivalence class generation}
\label{sec:background-motivation-and-formalism:formalism:error-equivalence-class-generation}

Under the STEPG representation, error analysis reduces to a straight-line program where the equivalence class $\epsilon_u$ of a vertex $u$ is computed as the bitwise XOR sum of its successors' classes:

\begin{equation} 
\label{eq:slp}
\epsilon_u = \bigoplus_{v \in N^+(u)} \epsilon_v
\end{equation}

The program is initialised by mapping the detector set $\mathcal{D}$ onto the leaf vertices. 
Each measurement $M$ is assigned to a leaf vertex, which serves as a boundary condition. 
For a detector $\delta \in \mathcal{D}$ defined by the parity of measurements $\{M_0, M_1, \dots\}$, we initialise the corresponding bit in the error equivalence classes:

\begin{equation}
\label{eq:init-slp}
\forall M \in \delta, \quad \epsilon_{\mathrm{leaf}(M)} \gets \epsilon_{\mathrm{leaf}(M)} \oplus \mathbf{e}_{\delta}
\end{equation}

where $\mathbf{e}_{\delta}$ is the elementary basis vector for detector $\delta$.

\subsubsection{Probability aggregation}
\label{sec:background-motivation-and-formalism:formalism:probability-aggregation}

Once every vertex has been assigned an equivalence class, the aggregate probability of each class $\epsilon$ is obtained by merging the probabilities $p_i$ of all physical errors mapped to it:

\begin{equation}
\label{eq:pmerge}
P(\epsilon) = \mathop{\boxplus}\limits_{i: E_i \mapsto \epsilon} p_i,
\end{equation}

where $p_a \boxplus p_b = p_a(1-p_b)+p_b(1-p_a)$.

\subsubsection{Beyond simple errors}
\label{sec:background-motivation-and-formalism:formalism:beyond-simple-errors}

\begin{figure}[htbp]
    \centering
    \includegraphics[width=\linewidth]{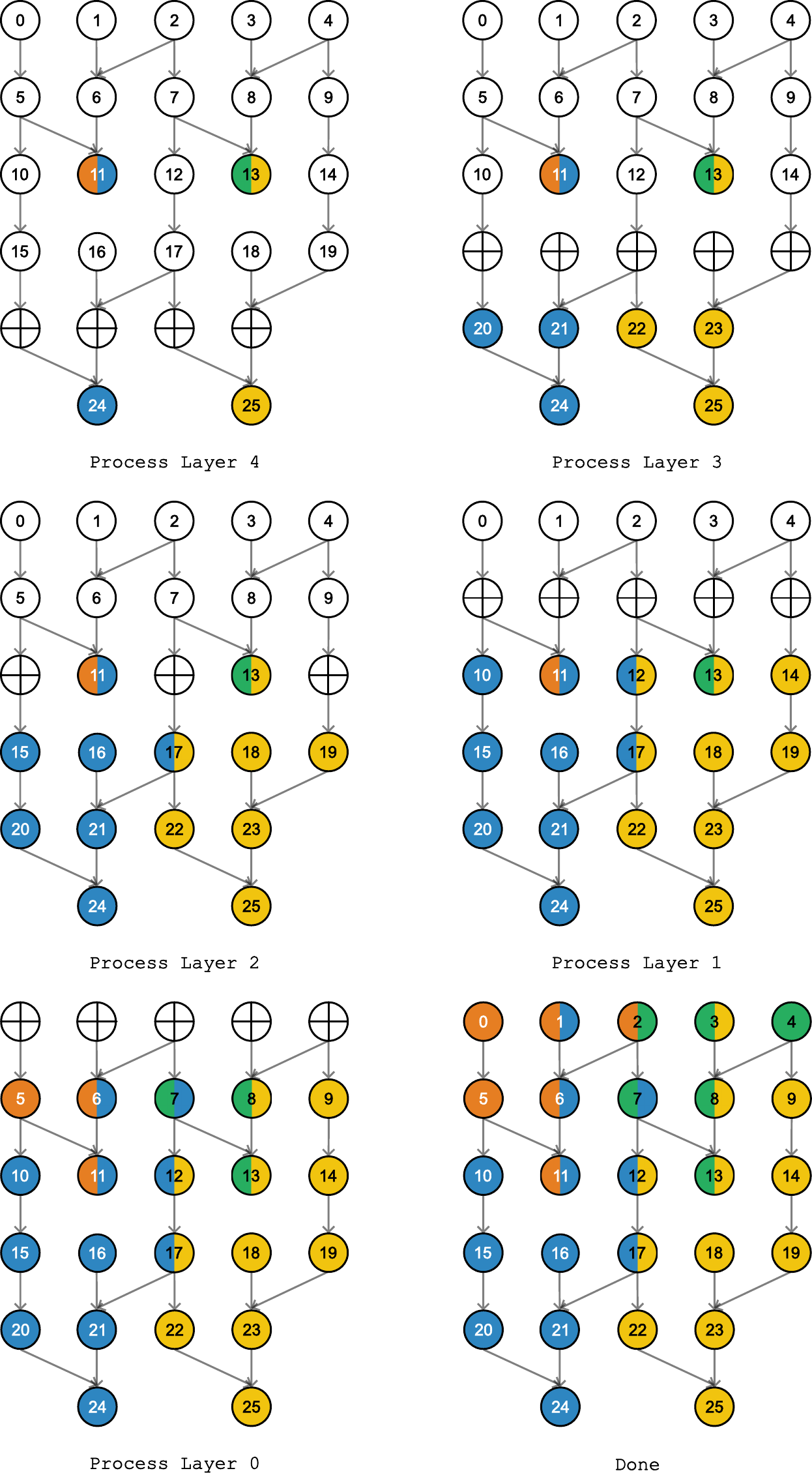}
    \caption{\textbf{A wave of XORs}.
    The straight-line program for error analysis (Eq.~\eqref{eq:slp}), implemented via a layer-synchronous backward traversal of the STEPG.
    The final layer is skipped (no out-edges); every preceding layer is processed in reverse, with one XOR per vertex executed in parallel across the layer.}
    \label{fig:stepg-wave}
\end{figure}

\figref{fig:stepg-wave} shows a STEPG for bit-flip ($X$) errors alone.
To include phase-flip ($Z$) errors, we use a symplectic representation~\cite{gottesman_stabilizer_1997}: the DAG is doubled so that each space-time location has $X$ and $Z$ vertices.
A correlated error such as $Y$ is then an additional source vertex whose equivalence class is the XOR of its $X$ and $Z$ components (Eq.~\eqref{eq:slp}).
We refer to such correlated errors as level~1 errors, and to higher-order combinations as level~2 (see~\tabref{tab:correlation-level}).

\section{Implementation}
\label{sec:implementation}

This section describes how GreenPeas turns the STEPG into a decoding hypergraph.
The pipeline consists of two stages: (1)~an error equivalence class generator that traverses the STEPG and assigns a class to each vertex; and (2)~a probability aggregator that deduplicates these classes and merges their probabilities into the final hypergraph.

\subsection{Error equivalence class generator}
\label{sec:compiler-design:eec-generator}

We implement the straight-line program for error analysis (Eq.~\eqref{eq:slp}) with a layer-synchronous backward traversal of the STEPG (Algorithm~\ref{alg:eec-generator}).
Within each layer, the $k=\alpha n$ vertices are independent and are processed in parallel, where $n$ is the number of qubits and $\alpha$ depends on the correlation level (Table~\ref{tab:correlation-level}).
This reduces the serial complexity from $\mathcal{O}(nl)$ to a parallel depth of $\mathcal{O}(l)$, where $l$ is the number of gate layers.

\SetKwFor{ParFor}{parallel for}{do}{end}

\begin{algorithm}
\caption{Error Equivalence Class Generator}
\label{alg:eec-generator}

\SetKwInOut{Input}{Input}

\Input{
  \begin{tabular}[t]{l @{\,:\,} l}
    $G$             & Space-time error propagation graph (STEPG) \\
    $l$             & Number of layers \\
    $k$             & Number of vertices per layer \\
    $\mathcal{E}$   & Error equivalence classes (modified in-place)
  \end{tabular}
}

\BlankLine

\For{$i \leftarrow l-2$ \KwTo $0$}{
  \BlankLine
  \ParFor{$j \leftarrow 0$ \KwTo $k-1$}{
    \BlankLine
    $u \leftarrow \text{GetIndex}(i, j)$
    \BlankLine
    $(v, w) \leftarrow \text{Unpack}(G[u])$
    \BlankLine
    $\mathcal{E}[u] \leftarrow \begin{cases} 
      \mathcal{E}[v] \oplus \mathcal{E}[w] & \text{if } v, w \neq -1 \\
      \mathcal{E}[v] & \text{if } v \neq -1, w = -1 \\
      \mathcal{E}[w] & \text{if } v = -1, w \neq -1 \\
      0 & \text{if } v = w = -1
    \end{cases}$
  }
}
\end{algorithm}

\vspace{0.5\baselineskip}
\begingroup
\setlength{\intextsep}{4pt}
\setlength{\abovecaptionskip}{4pt}
\setlength{\belowcaptionskip}{4pt}
\begin{table}[H]
\centering
\renewcommand{\arraystretch}{1.1}
\begin{tabularx}{\linewidth}{|c|X|c|}
\hline
\textbf{Correlation Level} & \textbf{Pauli Errors} & \textbf{Factor ($\alpha$)} \\ \hline
0 & $IX, IZ, XI, XX, ZI, ZZ$ & 3 \\ \hline
1 & $IY, XZ, YI, ZX$ & 5 \\ \hline
2 & $XY, YX, YY, YZ, ZY$ & 8 \\ \hline
\end{tabularx}
\caption{The $15$ non-trivial Pauli errors produced by a two-qubit gate classified into correlation levels ($L_0 \subset L_1 \subset L_2$).
         At each level, the factor $\alpha$---which determines the STEPG number of vertices per layer $k = \alpha n$---is given by half the size (rounded up) of the included error set at that level.}
\label{tab:correlation-level}
\end{table}
\endgroup

\subsubsection{Data layouts}
\label{sec:implementation:eec-generator:data-layouts}

We use GPU-oriented layouts for the STEPG and the equivalence class matrix.
The STEPG is stored as an array of 64-bit words whose high and low 32-bit halves hold the two successors of each vertex, so the $k$ threads of a layer can load their successor indices in one contiguous transaction.
Equivalence classes are stored in a bit-packed matrix with $W=\lceil|\mathcal{D}|/64\rceil$ 64-bit words per class.
A column-major layout keeps the $w$-th word of every class contiguous in memory, so the parallel XORs coalesce across the threads.

\subsubsection{Thread decomposition}
\label{sec:implementation:eec-generator:thread-decomposition}

With this representation, each $\oplus$ in Algorithm~\ref{alg:eec-generator} is a sequence of $W$ word-wise XOR operations.
When $W$ is large, we use a two-dimensional thread decomposition: one dimension indexes vertices in the layer and the other indexes words of the class bit vectors.
The kernel then scales with both the number of vertices and the number of detectors, improving GPU occupancy.

\subsubsection{Cache-friendliness}
\label{sec:implementation:eec-generator:cache-friendly}

The backward traversal order is favourable to the GPU cache hierarchy.
Successor indices $v,w$ for vertices in layer~$i$ refer only to vertices in the just-processed layer~$i+1$, so the bit arrays $\mathcal{E}[v]$ and $\mathcal{E}[w]$ have high temporal locality and the bitwise updates are typically served from cache rather than global memory.

\subsection{Probability aggregator}
\label{sec:implementation:probability-aggregator}

The aggregator groups vertices by equivalence class and merges their probabilities (Eq.~\eqref{eq:pmerge}) using a sort-and-reduce pipeline.
Since each class is a bit vector of $W>1$ 64-bit words, sorting these bit vectors is expensive.
We therefore hash each class to a fixed-width key and sort by that key.

We project each class to a 64-bit FNV-1a hash~\cite{estebanez_performance_2014}.
This hash is cheap to compute, and at the scales considered in this paper its key space is large enough that distinct classes are unlikely to collide and corrupt the decoding hypergraph, as could happen with 32-bit keys.

\subsection{Configurable correlation level}
\label{sec:implementation:configurable-correlation-level}

Prior work uses only the extremes of correlated error models: uncorrelated ($L_0$) or fully correlated ($L_2$).
GreenPeas instead exposes correlation levels $L_0\subset L_1\subset L_2$ that select which errors are included in the STEPG (Table~\ref{tab:correlation-level}).
Lower levels drop certain errors, producing sparser hypergraphs that compile and decode faster at the cost of logical accuracy.
Level $L_0$ keeps only pure $X$- and $Z$-type errors.
It excludes terms such as $Y$ or $XZ$ that can flip both $X$- and $Z$-type stabilisers.
In a $Z$-basis experiment, $X$-detectors should therefore be omitted (and conversely for an $X$-basis experiment).

\subsection{Buffer reuse}
\label{sec:implementation:buffer-reuse}

To avoid repeated dynamic memory allocation in online workflows, the GreenPeas compiler driver can be constructed from a maximum circuit that bounds the number of qubits, layers, measurements, and detectors.
Subsequent circuits that fit within those bounds then reuse the preallocated GPU buffers.
Listing~\ref{lst:basic-example} illustrates this pattern in the Python API, compiling each circuit to a Stim DEM.

\begin{lstlisting}[
  language=Python,
  basicstyle=\footnotesize\ttfamily,
  frame=single,
  breaklines=true,
  backgroundcolor=\color{white},
  float=htbp,
  captionpos=b,
  caption={Python API usage with GPU buffer reuse.},
  label={lst:basic-example}
]
import greenpeas as gp

# Construct max circuit.
max_circuit = get_max_circuit()

# Construct compiler driver from max circuit.
driver = gp.get_driver(max_circuit)

# Compile DEM for max circuit.
dem = driver.compile_detector_error_model()

# Reuse buffers for smaller circuits.
for c in iter_circuits():
    dem = driver.compile_detector_error_model(c)
\end{lstlisting}

\section{Results}
\label{sec:results}

We compare GreenPeas with Stim in terms of compilation time and of decoding accuracy and latency on the decoding hypergraphs they produce.
We perform standard $Z$-basis memory experiments under circuit-level noise, using both the surface code~\cite{kitaev_fault-tolerant_2003} and the bivariate bicycle code~\cite{bravyi_high-threshold_2024}.
From a compilation standpoint, these circuits serve as worst-case proxy workloads for round-by-round adaptive circuits when repeat block annotations are omitted: Stim's loop folding optimisation cannot then be applied.
This setup is chosen so that the reported speedups are accompanied by logical error rate estimates that can be cross-checked against prior results in the literature.
In~\secref{sec:case-study:results:circuit-level} we show that these speedups are even more pronounced in the true adaptive setting.

Since we target a per-cycle compilation latency of $\le\SI{1}{\ms}$ (dashed red line in the left panels of~\figref{fig:single-level}), we adopt a noise model inspired by hardware modalities with millisecond-scale cycle times, such as neutral-atom and trapped-ion systems.
For a physical error rate $p$, this noise model is parametrised as $p_1 = p_i = p/10$ and $p_2 = p_r = p_m = p$ (see~\secref{sec:background-motivation-and-formalism:background:noise-models:circuit-level} for definitions), following~\cite{berthusen_adaptive_2025}.

We evaluate decoding accuracy and latency using the Tesseract most-likely error decoder~\cite{beni_tesseract_2025} in the short-beam setting, applied to the decoding hypergraphs produced by GreenPeas and Stim.
For each memory experiment, we define the logical error rate $R$ as the fraction of samples where the decoder mispredicts the correction operator.
Since we use $d$ error correction cycles, the logical error rate per cycle

\begin{equation}
\label{eq:ler-per-cycle}
R_{\mathrm{per\,cycle}} = \frac12\Bigl(1-(1-2R)^{1/d}\Bigr).
\end{equation}

We measure GreenPeas compilation times on an NVIDIA Blackwell GPU, and Stim compilation times and Tesseract decoding latencies on an AMD EPYC CPU.
Per cycle times are obtained by normalising mean execution times by $d$.

\subsection{Subthreshold scaling}
\label{sec:results:subthreshold-scaling}

\newcolumntype{C}[1]{>{\centering\arraybackslash}p{#1}}

\begin{figure*}
    \centering
    \includegraphics[width=\textwidth]{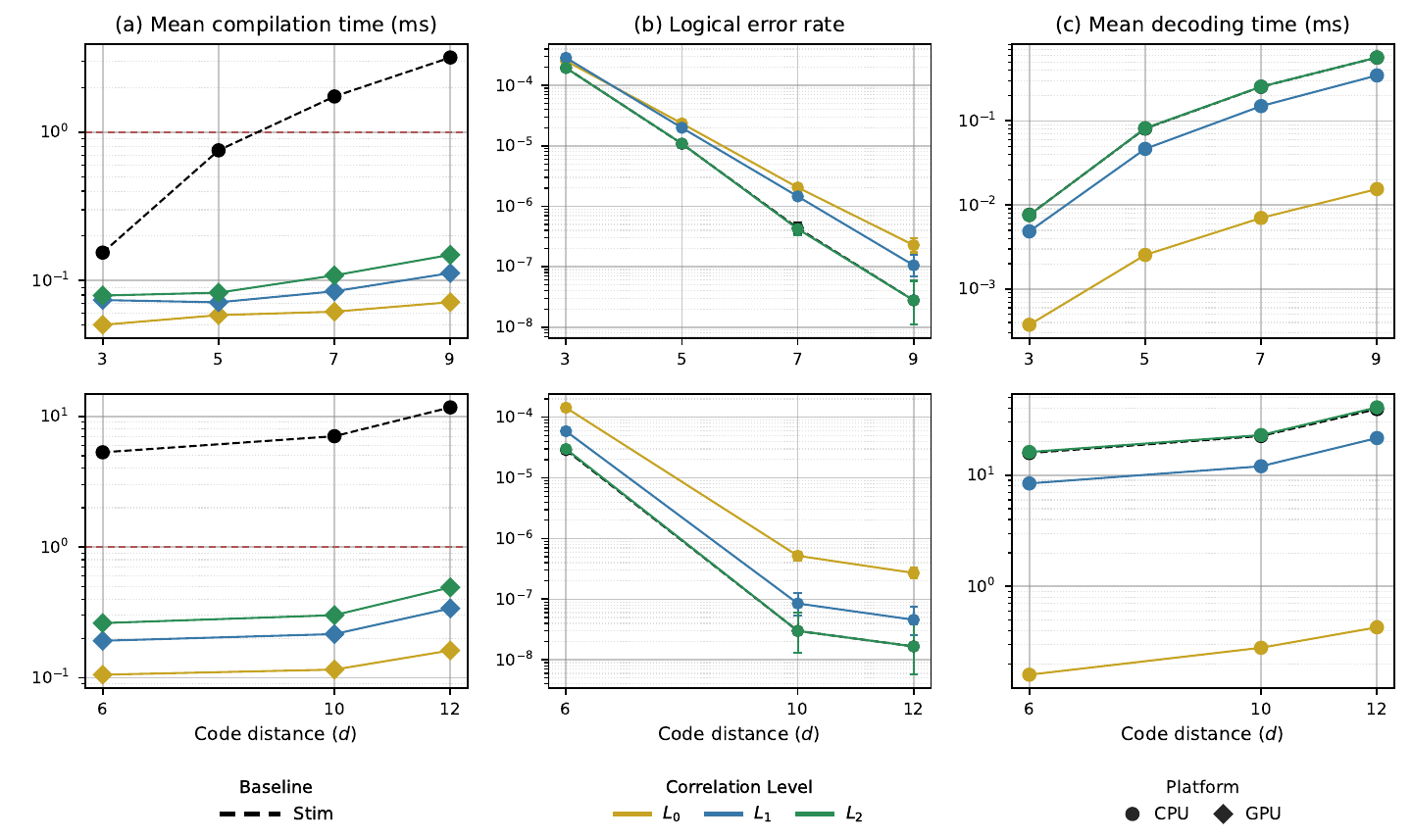}
    \caption{\textbf{Subthreshold scaling}.
    Top: surface codes; bottom: bivariate bicycle codes.
    $p = 0.1\%$ in all cases.
    All metrics are reported per cycle.
    \textbf{(a)} Mean compilation time, averaged over $10{,}000$ compiles of the same circuit.
    The dashed red line marks the $\SI{1}{\ms}$ target.
    \textbf{(b)} Logical error rate from $20$ million samples per data point; error bars are $90\%$ confidence intervals.
    The GreenPeas $L_2$ and Stim curves overlap, indicating no loss of decoding accuracy.
    \textbf{(c)} Mean decoding time, averaged over $10{,}000$ different syndromes.
    The GreenPeas $L_2$ and Stim curves again overlap, indicating no increase in decoding latency.
    Lower correlation levels trade accuracy for reduced compilation and decoding time.}
    \label{fig:single-level}
\end{figure*}

\begin{figure*}
    \centering
    \includegraphics[width=\textwidth]{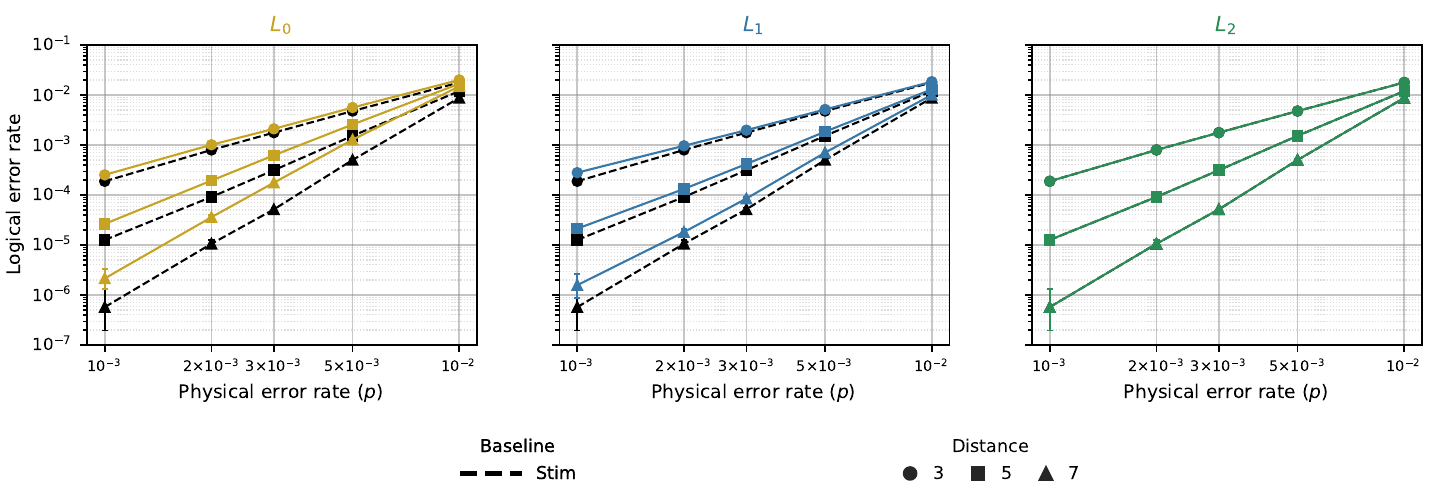}
    \caption{\textbf{Scaling with physical error rate}.
    Logical error rate per cycle versus~$p$ for surface codes with distances $d \in \{3,5,7\}$.
    Panels show GreenPeas~$L_0$, $L_1$, and $L_2$, each compared with Stim.
    GreenPeas~$L_2$ overlays Stim over the full plotted range; gaps between correlation levels shrink as $p$ approaches the surface code threshold.}
    \label{fig:ler-vs-p}
\end{figure*}

\begin{table}
  \centering
  \begin{tabular*}{\linewidth}{@{\extracolsep{\fill}}rccc@{\hspace{1.1em}}rccc@{}}
    \toprule
    \multicolumn{4}{c}{Surface code} &
    \multicolumn{4}{c}{Bivariate bicycle code} \\
    \cmidrule(r){1-4} \cmidrule(l){5-8}
    $d$ & $L_0$ & $L_1$ & $L_2$ &
    $d$ & $L_0$ & $L_1$ & $L_2$ \\
    \midrule
    3 & $3.06$ & $2.09$ & $1.95$ &
    6 & $50.24$ & $27.63$ & $20.24$ \\
    5 & $12.89$ & $10.57$ & $9.10$ &
    10 & $60.95$ & $32.61$ & $23.38$ \\
    7 & $28.19$ & $20.54$ & $16.05$ &
    12 & $72.27$ & $34.45$ & $23.79$ \\
    9 & $44.55$ & $28.36$ & $21.37$ &
      & & & \\
    \bottomrule
  \end{tabular*}
  \vspace{1ex}
  \caption{Compilation time speedups of GreenPeas relative to Stim for correlation levels $L_0$, $L_1$, and $L_2$ across surface and bivariate bicycle codes.
  The geometric mean speedups across all codes are $27.19\times$, $16.93\times$, and $13.16\times$, respectively.}
  \label{tab:single-level-speedups}
\end{table}

Figure~\ref{fig:single-level} shows how GreenPeas decoding hypergraphs scale with code distance at fixed $p = 0.1\%$.
For compilation time (\figref{fig:single-level}(a)), Stim and GreenPeas are directly comparable only at correlation level $L_2$, where the two tools produce equal hypergraphs up to minor differences in hyperedge weights arising from CPU versus GPU floating-point arithmetic.
At this level, GreenPeas yields a geometric-mean speedup of $13.2\times$ over Stim; speedups for $L_0$--$L_2$ are given in~\tabref{tab:single-level-speedups}.
For logical error rate and decoding time (Figs.~\ref{fig:single-level}(b) and (c)), GreenPeas $L_2$ and Stim remain equivalent, as shown by the overlapping curves.

The correlation level controls an accuracy-speed trade-off.
Lower levels reduce decoding latency at the cost of higher logical error rates; higher levels recover Stim-level accuracy with correspondingly higher latency.
As the code distance increases, the accuracy gap between different correlation levels widens, whereas the corresponding gap in decoding time remains comparatively stable.

\subsection{Scaling with physical error rate}
\label{sec:results:scaling-with-physical-error-rate}

Figure~\ref{fig:ler-vs-p} shows how the accuracy of GreenPeas decoding hypergraphs scales with~$p$ for surface code distances $d \in \{3,5,7\}$.
Each panel overlays Stim with a single GreenPeas correlation level ($L_0$, $L_1$, or $L_2$).
GreenPeas $L_2$ overlaps with Stim across the full range of $p$ and $d$, including near the code threshold of around~$1\%$.
By contrast, $L_0$ and $L_1$ incur an accuracy penalty that is most pronounced deep in the sub-threshold regime and at larger~$d$, and that shrinks as $p$ approaches threshold.
Thus the cheaper correlation levels are most competitive in the high-noise regime, while $L_2$ remains available when the highest fidelity level is required.

\section{Case study: adaptive SM circuits}
\label{sec:case-study}

Adaptive SM circuits are a natural fit for online hypergraph compilation: mid-circuit measurement outcomes change the circuit structure, so the hypergraph cannot be compiled once ahead of time. 
This section shows that GreenPeas enables circuit-level decoding of adaptive SM circuits for the [[4,2,2]]-concatenated surface code, improving both logical error rate and decoding latency relative to phenomenological decoding.

\subsection{Background}
\label{sec:case-study:background}

Fault-tolerance faces a fundamental challenge: measuring errors requires quantum gates, yet every gate introduces new noise.
Conventional SM circuits rely on a static set of measurements that can inadvertently overwhelm the code with the very errors it seeks to correct.
Adaptive SM circuits on the other hand employ a dynamic set of measurements via preliminary observations that isolate and target only probable error locations.

Berthusen et al.~\cite{berthusen_adaptive_2025} demonstrated the utility of adaptive SM circuits via a concatenated structure: an inner $[[4, 2, 2]]$ Iceberg code~\cite{self_protecting_2024} is combined  with an outer hypergraph product (HGP) code~\cite{tillich_quantum_2009}. 
The inner checks then filter the outer layer, restricting outer syndrome measurements to only those regions where the inner codes indicate the presence of errors, minimising the injection of noise from redundant, low-fidelity operations.

\subsubsection{$[[4,2,2]]$-concatenated surface code}
\label{sec:case-study:background:csc}

\begin{figure}
    \centering
    \includegraphics[width=0.85\linewidth]{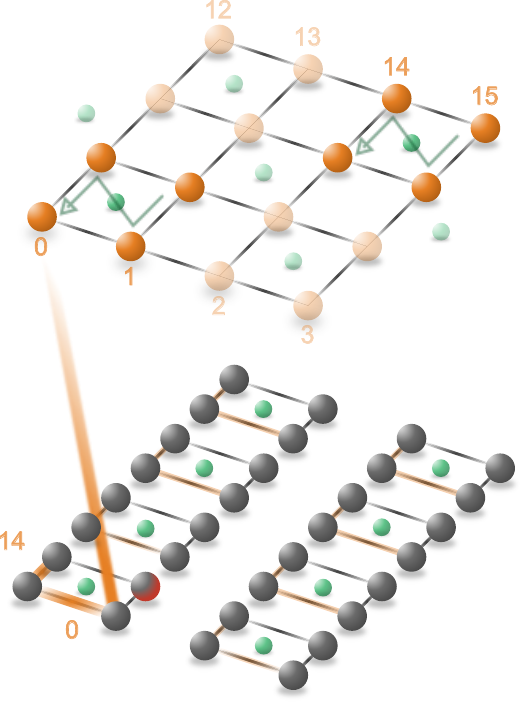}
    \caption{\textbf{$[[4,2,2]]$-concatenated surface code}.
    A $d = 4$ surface code (top) concatenated with an array of Iceberg codes (bottom).
    Data qubits in the surface code (orange) map to logical operators within the Iceberg blocks, e.g., data qubits 0 and 14 correspond to the first and second logical operators of the leading block, respectively.
    An initial error detection cycle measures the checks (green) of all Iceberg codes.
    In the subsequent error correction cycle, surface checks with support on data qubits in an excited Iceberg block, i.e., one containing an odd number of errors (red), are executed, e.g., the first and last surface checks are run since they support qubits 0 and 14, both of which belong to the excited Iceberg block.
    Other checks may be adaptively skipped, reducing the total gate count and logical error rate.}
    \label{fig:csc}
\end{figure}

This scheme is illustrated in~\figref{fig:csc} for an outer surface code.
To preserve fault-tolerance, data qubits supported by the same surface code check must be mapped to separate Iceberg blocks. 
This ensures that a single fault within a block cannot spread into a weight-2 error in the outer code, which would otherwise reduce its effective distance. 
Additionally, to ensure schedule-compatibility, the mapping must assign data qubits that share a time slot in the surface code’s schedule to separate Iceberg blocks, thereby preventing hardware-level gate conflicts.

\newpage

Finding a mapping that satisfies these constraints can be framed as a maximum matching problem on a conflict graph.
For regular lattices, however, it can be solved via geometric heuristics. 
Here, we employ the mapping shown in Fig.~\ref{fig:csc}. 
Assuming standard N- and Z-shaped CNOT sequences for X and Z checks, respectively, we pair each data qubit at coordinate $(r, c)$ in the surface code with a partner at

\begin{equation} 
\label{eq:iceberg-mapping}
(r', c') = (d - 1 - r, (c + 2) \bmod d)
\end{equation}

where $d$ is the distance of the surface code.

The reflection across the lattice ensures that paired qubits are supported by different surface code checks, while the cyclic shift ensures they occupy different time slots in its schedule, satisfying the two critical constraints.

\subsection{Simulation details}
\label{sec:case-study:simulation-details}

\subsubsection{Decoding hypergraph compilation}
\label{sec:case-study:simulation-details:decoding-hypergraph-compilation}

Adaptive SM circuits are difficult to simulate since decoding hypergraphs cannot be compiled before circuit simulation and decoding.
Each cycle, the set of active checks is chosen stochastically, so the number of potential hypergraphs grows exponentially with distance, making the memory cost of precompilation prohibitive for all but the smallest distances. 
We overcome this with GreenPeas: the hypergraph is compiled just-in-time, after circuit simulation and before decoding, retaining only the realised execution path in memory.

\subsubsection{Inner code decoding}
\label{sec:case-study:simulation-details:inner-code-decoding}

Following Section VI.A of~\cite{berthusen_adaptive_2025}, we decode the inner Iceberg code blocks prior to the outer code according to a heuristic recipe.
Specifically, for every detected X (Z) error, we apply a corresponding X (Z) Pauli correction to the shared qubit of the X (Z) logical operator.
However, to maintain high logical fidelity, the final decoding of the concatenated structure must integrate syndromes from both the inner and outer checks.

These heuristic Pauli corrections are applied each cycle based on the absolute outcomes of the current inner-check layer. 
The execution of the outer checks on the other hand is triggered by the difference syndrome: the XOR sum of the current and previous inner-check layers. 
This hybrid approach is consistent with the prior art, but employing difference syndromes to drive both correction and triggering logic may offer a more cohesive strategy. 
We defer a formal investigation of such a unified approach to future work.

\subsubsection{Refresh rate}
\label{sec:case-study:simulation-details:refresh-rate}

We execute all outer checks during the initial and final rounds of the experiment to mitigate boundary effects.
As established in Section VI.B of~\cite{berthusen_adaptive_2025}, it is also essential to apply all outer checks during intermediate rounds at a refresh rate $\gamma$ to prevent a build-up of logical errors in the inner code blocks.
In our $d$-round experiments, a refresh rate of $\gamma = d/2$ yielded the optimal logical fidelity and distance scaling; a full sweep of $\gamma$ is left to future work.

\subsubsection{Dynamic detector instantiation}
\label{sec:case-study:simulation-details:dynamic-detector-instantiation}

For each check, we track its current and previous measurement indices, and advance the current index only when the check is executed.
A detector is instantiated precisely when these indices differ, so the syndrome stays consistent with the adaptive circuit.
Relative to the static case, this yields fewer detectors on average and thus lower decoding latency.

\subsubsection{Choice of outer code}
\label{sec:case-study:simulation-details:choice-of-outer-code}

We use the surface code as the outer code to demonstrate circuit-level decoding enabled by online decoding hypergraph compilation.
Its low-weight checks make it a weaker fit for adaptive SM than codes with higher-weight checks, e.g., bivariate bicycle codes, but the theory for non-HGP constructions is less settled.
The surface code is therefore a convenient, well-understood vehicle for the purposes of this case study.

\subsubsection{Outer code decoding}
\label{sec:case-study:simulation-details:outer-code-decoding}

At levels $L_1$ and $L_2$, Tesseract decoding of the $[[4,2,2]]$-concatenated surface code---with or without adaptive measurements---was too slow for high-confidence logical error rate estimation.
We therefore use $L_0$ decoding hypergraphs for all concatenated instances; other beam settings or decoders may yet make $L_1$ or $L_2$ practical.

\subsection{Results}
\label{sec:case-study:results}

\subsubsection{Circuit-level decoding}
\label{sec:case-study:results:circuit-level}

\begin{figure*}
    \centering
    \includegraphics[width=\textwidth]{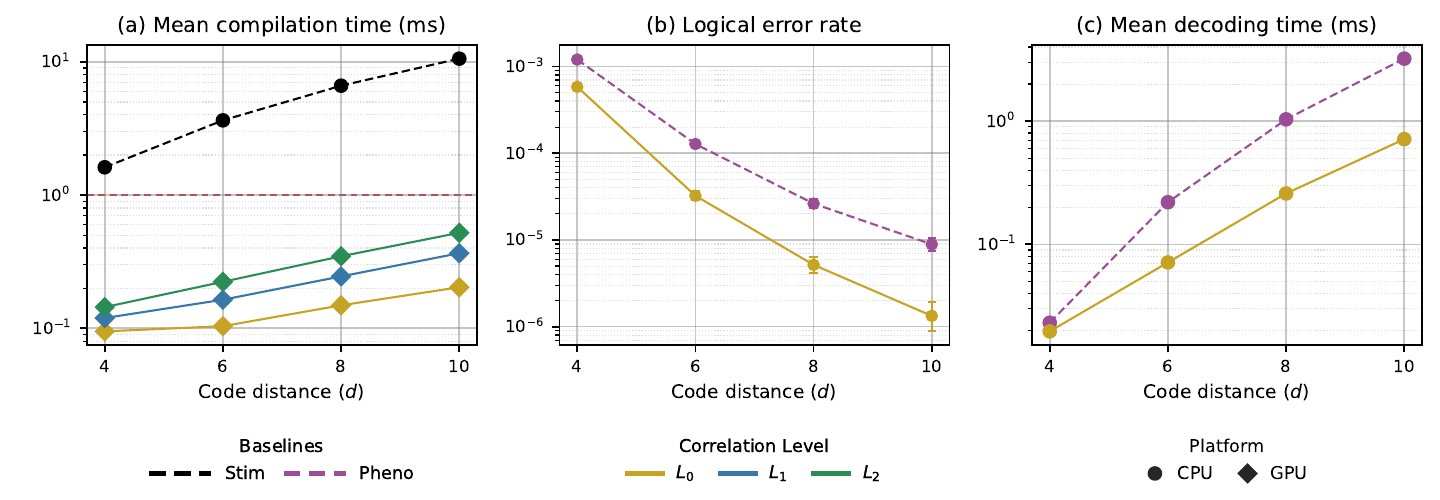}
    \caption{\textbf{Circuit-level decoding}.
    $[[4,2,2]]$-concatenated surface codes with adaptive SM ($p = 0.1\%$). 
    All metrics are reported per cycle.
    \textbf{(a)} Mean compilation time, averaged over $10{,}000$ adaptive circuit paths.
    The dashed red line marks the $\SI{1}{\ms}$ target.
    \textbf{(b)} Logical error rate from at least $1$ million samples; the phenomenological decoder is Stim-compiled.
    \textbf{(c)} Mean decoding time, averaged over $10{,}000$ different syndromes.
    Panels (b) and (c) report circuit-level $L_0$ only.
    Although GreenPeas compiles $L_1$ and $L_2$ hypergraphs in under a millisecond (panel (a)), decoding latency is too high for high-confidence logical error rate estimates.}
    \label{fig:multi-level}
\end{figure*}

\begin{table}
  \centering
  \begin{tabular*}{\linewidth}{@{\extracolsep{\fill}}rccc@{}}
    \toprule
    \multicolumn{4}{c}{Concatenated surface code} \\
    \cmidrule(lr){1-4}
    $d$ & $L_0$ & $L_1$ & $L_2$ \\
    \midrule
    4 & $17.04$ & $13.54$ & $11.21$ \\
    6 & $35.08$ & $22.21$ & $16.29$ \\
    8 & $44.51$ & $27.01$ & $19.07$ \\
    10 & $51.98$ & $28.88$ & $20.29$ \\
    \bottomrule
  \end{tabular*}
  \vspace{1ex}
  \caption{Compilation time speedups of GreenPeas relative to Stim for correlation levels $L_0$, $L_1$, and $L_2$ across $[[4,2,2]]$-concatenated surface codes with adaptive SM.
  The geometric mean speedups are $34.29\times$, $22.01\times$, and $16.31\times$, respectively.}
  \label{tab:multi-level-speedups}
\end{table}

\begin{figure}
    \centering
    \includegraphics[width=0.333\textwidth]{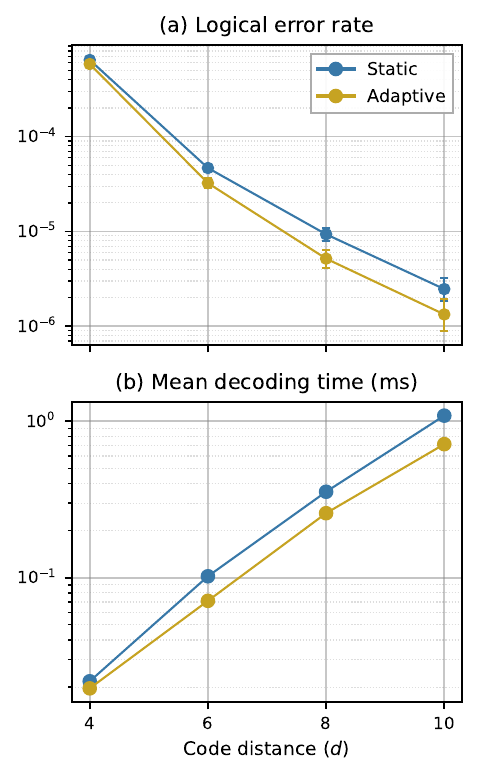}
    \caption{\textbf{Static vs. adaptive SM}.
    $[[4,2,2]]$-concatenated surface codes with $p = 0.1\%$.
    All data is reported per cycle.
    \textbf{(a)} Logical error rate from at least $1$ million samples.
    \textbf{(b)} Mean decoding time, averaged over $10{,}000$ different syndromes.
    Adaptive consistently outperforms static both in terms of logical error rate and mean decoding time, with the relative benefit increasing with code distance.}
    \label{fig:static-vs-adaptive}
\end{figure}

Prior work on adaptive SM circuits relied on phenomenological decoding~\cite{berthusen_adaptive_2025}, owing to the prohibitive cost of compiling a circuit-level hypergraph on every shot.
As shown in Fig.~\ref{fig:multi-level}(a), GreenPeas reduces per-cycle compilation latency for such circuits to below one millisecond, making circuit-level decoding practical in these experiments.
In particular, GreenPeas $L_2$ yields a geometric-mean speedup of $16.3\times$ over Stim; speedups for $L_0$--$L_2$ are reported in Table~\ref{tab:multi-level-speedups}.
Figures~\ref{fig:multi-level}(b) and~(c) compare circuit-level $L_0$ decoding with a Stim-compiled phenomenological decoder.
At outer-code distance~$10$, circuit-level decoding achieves a $6.7\times$ lower logical error rate and a $4.5\times$ lower decoding latency.

\subsubsection{Static vs. adaptive SM}
\label{sec:case-study:results:static-vs-adaptive}

Figure~\ref{fig:static-vs-adaptive} compares static and adaptive SM on the $[[4,2,2]]$-concatenated surface code.
The relative logical error rate advantage of adaptive SM grows with code distance~$d$: at $d = 10$, the adaptive scheme nearly halves the logical error rate relative to the static case.
This growth with~$d$ is largely expected: our memory experiments use~$d$ error-correction rounds, so larger distances include more adaptive rounds and accumulate a greater advantage over static SM.
The adaptive strategy also reduces decoding time, since fewer detectors are active on average; at $d = 10$ we observe around a $35\%$ reduction in decoding latency.

\subsubsection{Comparison with prior art}
\label{sec:case-study:results:prior-art}

A direct comparison with~\cite{berthusen_adaptive_2025} is precluded by the use of different outer codes.
Both works perform circuit-level simulations; however,~\cite{berthusen_adaptive_2025} decodes with a phenomenological decoder consisting of a combination of BP-OSD~\cite{hillmann_localized_2025} and BP-LSD~\cite{roffe_decoding_2020}, whereas we use a full circuit-level decoder via GreenPeas and Tesseract.
At the same physical error rate $p = 0.1\%$, our results show clear exponential suppression of the logical error rate with increasing distance, consistent with sub-threshold operation.

\section{Discussion}
\label{sec:discussion}

\subsection{Related work: deq}

Microsoft's deq~\cite{Microsoft_QDK_EC} is the closest related work to GreenPeas.
Designed for dynamic logical circuits, it organises execution around gadget structures whose decoding hypergraphs are compiled via a hybrid offline/online workflow: gadget types are built offline, then stitched together as logical instructions stream in at runtime.
Since error analysis is performed in the offline stage, deq is less naturally suited to physical-level branching, where mid-circuit measurement outcomes change the physical circuit itself, making it difficult to define a fixed gadget library.
GreenPeas and deq therefore address complementary layers in the adaptive-circuit runtime stack.

\subsection{Positioning relative to Stim}
\label{sec:discussion:positioning}

While this paper evaluates GreenPeas against Stim, the two are fundamentally complementary.
The latter provides a comprehensive ecosystem for the simulation and analysis of Clifford circuits. In contrast, the former is a specialised toolchain designed for the specific bottleneck of compiling decoding hypergraphs. 
This specialisation is particularly critical for adaptive circuits, where compilation must occur at runtime and meet strict latency requirements. 
For static or offline use cases, Stim remains the industry standard.

\subsection{CPU vs. GPU comparisons}
\label{sec:discussion:cpu-vs-gpu}

Although Stim's error analysis pipeline leverages only a single CPU core, it remains the state-of-the-art reference for decoding hypergraph compilation. 
A multicore CPU version of GreenPeas would be fundamentally mismatched, since its algorithm is designed for massive thread concurrency where each thread performs a small unit of work. 
Such fine-grained parallelism would lead to prohibitive scheduling overhead on a multicore CPU, whereas it is precisely what the GPU architecture is designed to support.

\subsection{Open challenges in adaptive circuit simulation}
\label{sec:discussion:open-challenges}

Low-latency decoding hypergraph compilation removes one bottleneck, but large-scale adaptive circuit simulations remain costly, especially at physical error rates $\le 0.1\%$, where tight error bars can demand $10^6$--$10^9$ Monte Carlo trials.
Two further challenges stand out: the cost of error sampling and the latency of decoder initialisation.

\subsubsection{Error sampling}
\label{sec:discussion:open-challenges:error-sampling}

Because adaptive circuits branch on mid-circuit measurements, simulations must use Stim's tableau simulator, which is orders of magnitude slower than its frame simulator.
This bottleneck is especially acute for high-branching-factor circuits such as those in~\secref{sec:case-study}.
Large-scale exploration of these circuits will likely need simulation methods that treat measurement-driven branching as a first-class primitive rather than a high-overhead exception.

\subsubsection{Decoder initialisation}
\label{sec:discussion:open-challenges:decoder-initialisation}

Even with fast hypergraph compilation, many decoders still pay a large initialisation cost before they can decode.
Some decoding accelerators, such as the Local Clustering Decoder~\cite{ziad_local_2025}, support runtime reconfiguration and are thus attractive for adaptive workflows, but they are typically specialised to restricted algorithms and graph models.
Developing low-overhead decoder initialisation---or amortising it across adaptive branches---remains an important open problem.

\subsection{Towards microsecond timescales}
\label{sec:discussion:towards-microsecond-timescales}

GreenPeas targets platforms with millisecond-scale cycle times, such as neutral-atom and trapped-ion systems.
These architectures natively support the long-range connectivity used by high-rate qLDPC codes and are a leading route to utility-scale quantum computing~\cite{webster_pinnacle_2026,cain_shors_2026,babbush_securing_2026,mundada_heterogeneous_2026}.
Solid-state platforms such as superconducting qubits, by contrast, are attractive for their microsecond-scale cycles, so we briefly discuss how our approach might extend to that regime.

\newpage

The error equivalence class generator (Algorithm~\ref{alg:eec-generator}) is highly parallel with a regular dependency structure, so it is a natural candidate for specialised acceleration, such as on an FPGA or ASIC.
By contrast, the probability aggregator is a sort-and-reduce pipeline that maps well onto GPUs.
This suggests a hybrid design, where the generator is offloaded to a spatial accelerator whilst the aggregator remains on the GPU, as one path to microsecond-scale devices.

\section{Conclusion}
\label{sec:conclusion}

Adaptive circuits are expected to play a central role in fault-tolerant quantum computing: they appear in many logical operations and can also reduce the error-correction overhead.
As shown here, adaptive syndrome measurements reduce logical errors and decoding latency by strategically omitting redundant checks.
Yet circuit-level decoding still depends on decoding hypergraph compilation, and existing offline approaches scale poorly when applied to adaptive circuits with large branching factors.

To this end, we presented GreenPeas, an online, just-in-time compiler for decoding hypergraphs.
On proxy workloads, it achieves a geometric mean speedup of $13.2\times$ over Stim.
This advantage carries over to the adaptive regime, where fast online compilation enables circuit-level decoding of $[[4,2,2]]$-concatenated surface codes with adaptive syndrome measurements.
Relative to prior phenomenological methods at outer-code distance~$10$, this reduces the logical error rate by $6.7\times$ and decoding latency by $4.5\times$.

\section{Author contributions}

Abbas B. Ziad conceived of the project, designed the core algorithms, implemented the open-source code, performed and analysed the experiments, and wrote the paper.
Jubo Xu ran a set of BP-OSD experiments to verify a hypothesis about correlation levels.
Hongxiang Fan supervised the project.
All authors contributed to the paper editing.
Generative AI was used to assist with routine coding tasks, such as scripting and plotting, and for light editing of article text.
The authors reviewed and take full responsibility for all content.

\label{sec:conclusion}

\clearpage

\bibliographystyle{ACM-Reference-Format}
\bibliography{references}

\end{document}